\documentclass[prl,twocolumn,superscriptaddress,floatfix,nopacs]{revtex4-1}
\usepackage{graphicx,amsfonts,amssymb,amsmath,hyperref,enumerate}
\usepackage{lmodern,textcomp} 
\usepackage{bbm}
\usepackage{ulem}
\usepackage{multirow}

\newif\ifhyper
\hypertrue
\ifhyper
\hypersetup{
   citecolor = {green},
   colorlinks = {true}, 
   urlcolor = {blue} 
}

\newcommand{\bra}[1]{{\langle #1 |}}
\newcommand{\ket}[1]{{| #1 \rangle}}

\begin{document} 

\title{Quantum portfolio value forecasting}

\author{Cristina Sanz-Fern\'andez}
\affiliation{Multiverse Computing, Paseo de Miram\'on 170, E-20014 San Sebasti\'an, Spain}

\author{Rodrigo Hern\'andez}
\affiliation{Multiverse Computing, Paseo de Miram\'on 170, E-20014 San Sebasti\'an, Spain}

\author{Christian D. Marciniak}
\affiliation{Institut f{\"u}r Experimentalphysik, 6020 Innsbruck, Austria}

\author{Ivan Pogorelov}
\affiliation{Institut f{\"u}r Experimentalphysik, 6020 Innsbruck, Austria}

\author{Thomas Monz}
\affiliation{Institut f{\"u}r Experimentalphysik, 6020 Innsbruck, Austria}
\affiliation{AQT, Technikerstr. 17, 6020 Innsbruck}

\author{Francesco Benfenati}
\affiliation{Multiverse Computing, Paseo de Miram\'on 170, E-20014 San Sebasti\'an, Spain}

\author{Samuel Mugel}
\affiliation{Multiverse Computing,  192 Spadina Ave, Suite 319 Toronto M5T 2C2, Canada}

\author{Rom\'an Or\'us}
\affiliation{Multiverse Computing, Paseo de Miram\'on 170, E-20014 San Sebasti\'an, Spain}
\affiliation{Ikerbasque Foundation for Science, Maria Diaz de Haro 3, E-48013 Bilbao, Spain}
\affiliation{Donostia International Physics Center, Paseo Manuel de Lardizabal 4, E-20018 San Sebasti\'an, Spain}

\begin{abstract}
We present an algorithm which efficiently estimates the intrinsic long-term value of a portfolio of assets on a quantum computer. The method relies on quantum amplitude estimation to estimate the mean of a novel implementation of the Gordon-Shapiro formula. The choice of loading and readout algorithms makes it possible to price a five-asset portfolio on present day quantum computers, a feat which has not been realised using quantum computing to date. We compare results from two available trapped ion quantum computers. Our results are consistent with classical benchmarks, but result in smaller statistical errors for the same computational cost.
\end{abstract}

\maketitle

\emph{Introduction.-} Monte Carlo methods are a hallmark of modern computing. These algorithms are essential for modelling random processes, as well as numerical integration and various optimization algorithms. They are ubiquitous in finance, where they are used to estimate risk and uncertainty. They are applied in fields which are intrinsically uncertain, such as option pricing, stock management, and default risk estimation \cite{wilmott2013}. As such, Monte Carlo simulations also typically represent the largest computational cost to financial institutions. Using these methods one can estimate risk up to an arbitrary precision, at the cost of a large number of runs. Financial institutions such as banks need to estimate risk with very high accuracies in order to minimize losses, and it is not uncommon for them to run simulations which take upwards of 24 hours in the best case.

In parallel to the above, quantum computing is known to accelerate many computational tasks. Some of these can be proven to not be achievable through any classical means (such as the square-root speed-up when searching in unstructured databases \cite{10.1145/237814.237866}). In fact, quantum computing has already found many applications in finance -- for a detailed review, see Ref.~\cite{Orus2018}. In this context, Montanaro \cite{Montanaro2015, An2020} was the first to suggest using quantum amplitude estimation to accelerate Monte Carlo methods. He showed that quantum computers can run Monte Carlo simulations to the same accuracy as classical computers, but reducing quadratically the number of samples, thus resulting in overall faster  calculations. Variations of this method have been successfully applied to toy academic models in finance in Refs.~\cite{Rebentrost2018, Woerner2019, Ramos-Calderer2019, Stamatopoulos2019}. 

In this paper, we develop a quantum method to estimate the intrinsic long-term value of a portfolio of assets, and implement it with real-life data. This is a core open problem in finance, which relies on an accurate forecasting of risk and market predictions. Accurate intrinsic long-term value estimation is crucial to portfolio optimization applications, which has been heavily investigated in the context of quantum computing \cite{Mugel2020a, Mugel2020b, Mugel2020c, Rosenberg2016, Elsokkary2017a, Grant2020, Cohen2020,Palmer2021}. Moreover, we implement the method in two different quantum computing architectures based on trapped ions: one provided by IonQ \footnote{See \href{https://ionq.com/}{https://ionq.com/}.}, and one described in Ref.~\cite{PRXQuantum.2.020343} (AQTION project). The results provide a use case of both quantum platforms when applied to a real financial problem. We choose to work with trapped ions because they provide a natural all-to-all connectivity among the qubits. Importantly, this property makes it simpler to implement the quantum circuit required in this work, in turn avoiding errors in, e.g., arbitrary 2-qubit gates between non-neighbouring qubits.
\bigskip 

\emph{Mathematical model.-} In general, an asset's intrinsic long-term value is the present value of all future cash flows \cite{williams1938theory}. If a portfolio's value is growing, its intrinsic value is given by the Gordon-Shapiro model \cite{DAmico2020}. It is possible to price assets based on their dividend yield at all times in the future. In practice, this can cause issues, as the intrinsic price diverges in certain parameter regimes as time goes to infinity. In this paper, we consider a truncated Gordon-Shapiro model, which only considers the assets' growth over the first two years, followed by a period at maturity. For more in-depth information, see Refs.~\cite{williams1938theory, DAmico2020}. 

Typically, the Gordon-Shapiro model prices assets based on the dividend yield. These do not always provide an accurate estimation of the assets' value, particularly for immature companies, or companies that have a culture of distributing little or no dividends. Thus, we prefer to evaluate asset value based on the asset's earnings per share. Our modified Gordon-Shapiro model is therefore given by:
\begin{equation}
\label{eq:gordon_shapiro_full}
    V = \sum_{j=0}^{n_{\rm{a}}-1} \omega_j v_j ,
\end{equation}
where $V$ is the total value of the portfolio, $j$ indicates the asset's index,  $n_{\rm{a}}$ the total number of assets, $\omega_j$ is the weight of asset $j$ (normalized to the total amount of investment), and $v_j$ is the single asset intrinsic value. This last quantity is given in the model by:
\begin{equation}
\label{eq:gordon_shapiro}
    v_j = \frac{E_{j,1}}{1+r_{j,1}}
    + \frac{E_{j,2}}{(1+r_{j,1})(1+r_{j,2})} \left(
    1 + \frac{1+g_j}{r_{j,\infty}-g_j}
    \right).
\end{equation}
In the equation above, $E_{j,1}$ and $E_{j,2}$ are the earnings per share (EPS) of asset $j$ respectively for the first and second year. The first-year, second-year, and long-term discount rates are given by $r_{j,1}$, $r_{j,2}$, and $r_{j,\infty}$, respectively. The EPS long-term growth rate of asset $j$ is given by $g_j$. 

In the following, we will assume that the short-term variables $E_{j,1}$, $r_{j,1}$, and $r_{j,2}$, as well as the growth rates $g_j$ can be forecasted to reasonable accuracy, and that the long-term variables $E_{j,2}$ and $r_{j, \infty}$ are harder to accurately predict, so that we  include small relative stochastic shifts $\delta_{j,E}$ and $\delta_{r}$. In general, all variables from Eq.~\eqref{eq:gordon_shapiro} could be described as stochastic variables themselves. In our implementation, however, we are motivated to approximate the short-term variables by their mean values as these only contribute weakly to Eq.~\eqref{eq:gordon_shapiro}.

Following the above, the long-term variables can be expressed as:
\begin{eqnarray}
\label{eq:stochastic_EPS}
E_{j,2} & = & \bar{E}_{j,2} (1 + \delta_{j, E}) = E_{j,1} (1 + g_j) (1 + \delta_{j, E}), \\
\label{eq:stochastic_r}
r_{j, \infty} & = & \bar{r}_{\infty} (1 + \delta_{r}) + \nu_j,
\end{eqnarray}
where $\bar{E}_{j,2}$ is the mean of $E_{j,2}$, $\bar{r}_{\infty}$ is the cost of capital for a risk-free asset, and $\nu_j$ is a fixed risk premium. In Eq.~\eqref{eq:stochastic_EPS}, we used the share's growth rate to estimate the mean EPS in the second year. Moreover, $\delta_{j, E}$ and $\delta_{r}$ are stochastic variables following a probability distribution. 

Including the above in Eq.~\eqref{eq:gordon_shapiro_full}, we see that the intrinsic value $V$ is itself a stochastic quantity that follows a probability distribution. In the following, we find it useful to discretise its value over $N$ states, each with value $V_i$ and probability $p(i)$. Our goal is to compute the mean intrinsic value:
\begin{equation}\label{eq:mean_value_Pport}
    \mathbb{E} [V] = \sum_{i=0}^{N-1} p(i) V_i,
\end{equation}
where $i$ is the discretization index. The aim of our quantum algorithm is, precisely, to estimate such expectation value as efficiently and accurately as possible. 

\bigskip 

\emph{Quantum algorithm.-} We discuss here briefly the main features of our quantum implementation, and leave the more in-depth technical details for the Supplementary Material. Let us start by saying that we encode all the possible discretised intrinsic values $V_i$ with probability $p(i)$  in quantum states, and then implement an algorithm which allows us to efficiently estimate the mean intrinsic value provided in  Eq.~\eqref{eq:mean_value_Pport} by using ``quantum-enhanced" Monte Carlo. The main steps involved are:
\begin{enumerate}
\item{{\it State preparation:} encode the distribution of the stochastic variables into a quantum state.}
\item{{\it Manipulation and readout:} manipulate the quantum state in such a way that we can efficiently estimate the mean of the probability distribution with the smallest possible error.}
\end{enumerate}

Concerning state preparation, the first step requires the definition of a Grover-like loading operator $\mathcal{A} = \mathcal{W}(\mathcal{P} \otimes \mathbbm{1})$, as required for quantum amplitude estimation. On the one hand, the operator $\mathcal{P}$ encodes a known probability distribution to $n$ qubits:
\begin{equation}
\label{eq:P-operator}
\mathcal{P}\ket{0}_n = \sum_{i=0}^{N-1} \sqrt{p(i)} \ket{i},
\end{equation}
where $N = 2^n$ is the total number of states and $\sum_{i=0}^{N-1} p(i) = 1$. We discuss the choice of this operator in the Supplementary Material. On the other hand, $\mathcal{W}$ is designed in such a way that it allows the encoding of a function $f(\cdot)$ in an ancilla qubit, in such a way that this enconding gets entangled with the probability distribution as follows:
\begin{equation}
\label{eq:qaa-operator}
\begin{split}
    \ket{\psi}_{n+1} &\equiv \mathcal{A}\ket{0}_{n+1} \\
    &= \sum_{i=0}^{N-1} \sqrt{p(i)} \ket{i} \left( \sqrt{1 - f(i)} \ket{0} + \sqrt{f(i)} \ket{1} \right) \; ,
\end{split}
\end{equation}
with $f(i) \in [0, 1]$. Importantly, this implies 
\begin{equation}
\label{eq:prob_1_state}
    _{n+1}\bra{\psi} (\mathbbm{1}_n \otimes \ket{1}\bra{1}) \ket{\psi}_{n+1} = \sum_{i=0}^{N-1} p(i) f(i) \; ,
\end{equation}
so that the expectation value of measuring $\ket{1}$ in the ancilla qubit is, in fact, the mean value of $f(\cdot)$. This is exactly how we are going to compute the mean intrinsic value of Eq.~\eqref{eq:mean_value_Pport}.

\begin{table}
\centering
\begin{tabular}{| l| | c c c c c |}
	\hline
	~\textbf{Index}~ & ~\textbf{SXXP}~ & \textbf{SPX}~ & \textbf{NKY}~ & \textbf{MXEF}~ & \textbf{EPRA}~ \\
	\hline
	~Holdings $\omega_j$~ & 2.75 & 19.5 & 0.04 & 0.89 & 0.59 \\
	\hline
\end{tabular}
\caption{Portfolio composition. Top row: index name. Bottom row: asset holdings $\omega_j$.}
\label{tab:holdings}
\end{table}

In general, choosing $\mathcal{W}$ such that we can effectively attach $V_i \to f(i)$ requires a significant overhead in circuit depth or ancilla qubits \cite{Haner2018, Mitarai2019}. Instead, we use the algorithm described in Ref.~\cite{Woerner2019}, which proposes an efficient way of implementing $\mathcal{W}$ via controlled-Y rotations, provided $V_i (\delta_{E}, \delta_r)$ is a polynomial function in the stochastic variables. The latter is achieved using the Taylor expansion of $V_i$, for which we assume here just linear order given that $\delta_{E}, \delta_r \ll 1$. We furthermore rescale $V_i$ from $0$ to $1$ as $\tilde{V}_i$, for implementation purposes. The details of the Taylor expansion and rescaling can be found in the Supplementary Material. 

As for the readout, we have constructed a controlled rotation $\mathcal{W}$ which maps the expectation value of the intrinsic price to the amplitude of an ancilla qubit as in Eq.~\eqref{eq:qaa-operator}. This can be done efficiently using, e.g., the quantum implementation of the reversible classical circuit for the function. Then, we estimate this amplitude efficiently using Quantum Amplitude Amplification (QAA) \cite{Brassard2002, Montanaro2015}. Usually this is done using approaches based on  phase estimation \cite{kitaev1995quantum}. However, here we prefer the Maximum Likelihood Estimation (MLE) method from Ref.~\cite{Suzuki2020}, which has low circuit depth, does not rely on expensive controlled rotations, and maintains the quadratic speedup. This method is described in the Supplementary Material. 

\bigskip 

\emph{Results.-} As a proof of concept, we study a portfolio of five assets, with each one of them being an index. We invest 1000 euros in each asset, with a full value of 5000 euros, bought at market value on 2021-04-25. The corresponding number of shares being held is shown in Table~\ref{tab:holdings}. Forecasted EPS, growth rates, and discount rates for each asset were purchased from a market data provider. All the input data used is summarised in the Supplementary Material.

\begin{figure}
  \centering
      \includegraphics[width=0.49\textwidth]{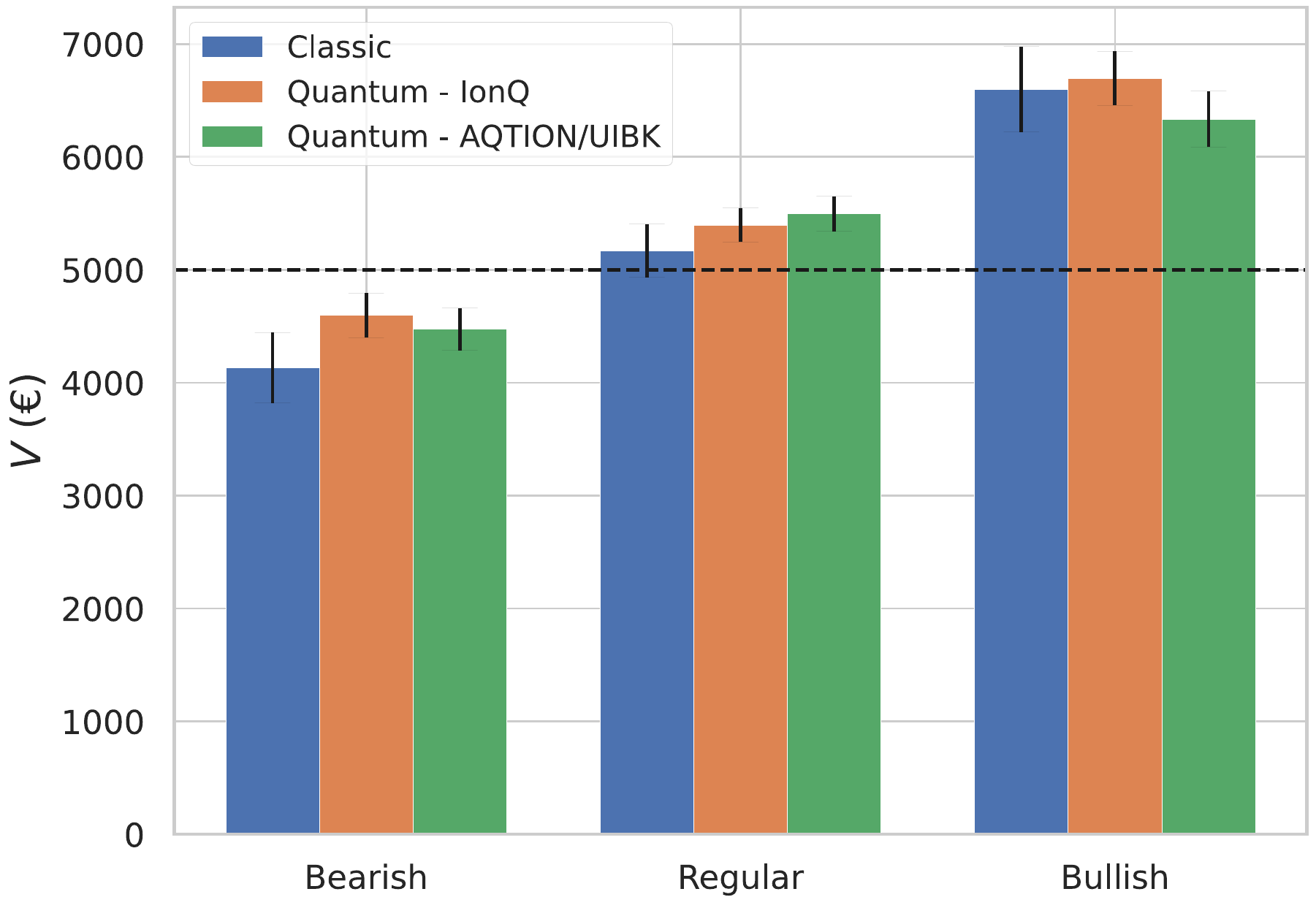}
  \caption{[Color online] Intrinsic value of the portfolio in Table~\ref{tab:holdings}. The intrinsic long-term value was forecasted using classical Monte Carlo (blue) and ``quantum-enhanced" Monte Carlo, as evaluated using IonQ's (orange) and AQTION's (green) QPUs. We studied three market trends: bearish (left), stable (center), and bullish (right). These results correspond to $4000$ queries to operator $\mathcal{A}$ (i.e., $4000$ calls to the function to be estimated). The error bars are the estimation error. The horizontal dashed line represents the portfolio's actual market value. We can see that the quantum result overlaps with the classical one in the three scenarios, meaning that the outcome of the quantum calculations are compatible with the classical ones, which we take as baseline. In fact, we observe that predictions are highly correlated for every run of the quantum circuit (not shown). One can also see that the estimation error for the quantum approach is smaller than the classical one.} 
  \label{fig:fair_value}
\end{figure}
\begin{table}
\centering
\begin{tabular}{| c || c c c |}
	\hline
	~\textbf{Market}~ & ~\textbf{Classical}~ & ~\textbf{IonQ}~ & ~\textbf{AQTION}~  \\
	\hline
	~Bearish~ & $7.6\%$ & $4.3\%$ & $4.2\%$  \\
	\hline
	~Regular~ & $4.6\%$ & $2.8\%$ & $2.8\%$  \\
	\hline
	~Bullish~ & $5.7\%$ & $3.6\%$ & $3.9\%$  \\
	\hline
\end{tabular}
\caption{Value of the estimation errors in Fig.~\ref{fig:fair_value}.}
\label{tab:markets}
\end{table}

The forecasted intrinsic values of the portfolio are presented in Fig.~\ref{fig:fair_value}. Technical details on the algorithm and the error calculation are in the Supplementary Material. We compare the results obtained using ``quantum-enhanced" Monte Carlo to those obtained using classical Monte Carlo for an equal amount of samples. Our circuits are run on two quantum hardware platforms based on trapped ions: IonQ and AQTION quantum processing units (QPUs). In order to evaluate the resilience of our model, we compare the results from three different market trends: \textit{bearish} or receding, regular or stable, and \textit{bullish} or rising. These behaviours are obtained by tuning the volatility and EPS long-term growth ($g_j$). We assume bearish and bullish markets to be more volatile than the stable ones (see the size of the error bars in Fig.~\ref{fig:fair_value}), and set $g_j \lesssim 0$ for bearish market, $g_j \gtrsim 0$ for stable markets, and $g_j > 0$ for bullish markets.

We see that, when the market is stable or bullish, the portfolio's intrinsic value exceeds the portfolio's market value. This implies that this portfolio is a sound investment, i.e: the rewards from buying this portfolio justify the risks. In this example, this is not the case when the market behaves bearishly, as the market overestimates the portfolio's value. Additionally, the error bars in Fig.~\ref{fig:fair_value} (exact values in Table~\ref{tab:markets}) show the confidence interval of the intrinsic price. These are larger in the classical case than the quantum case. This is because, by virtue of Chebyshev's inequality \footnote{Let $X$ be a random variable with a finite expected value $\mu$ and finite non-zero variance $\sigma^2$. Then, for any real number $k > 0$, one has that $Pr(| X - \mu | \ge k \sigma ) \le 1/k^2$ \cite{Cheby}.}, the computational cost of classically calculating the intrinsic value within distance $\epsilon$ of its actual value becomes prohibitively large for small $\epsilon$. This inequality does not hold in the quantum case, and the quantum implementation is more efficient by a quadratic factor. 

\begin{figure}
  \centering
      \includegraphics[width=0.48\textwidth]{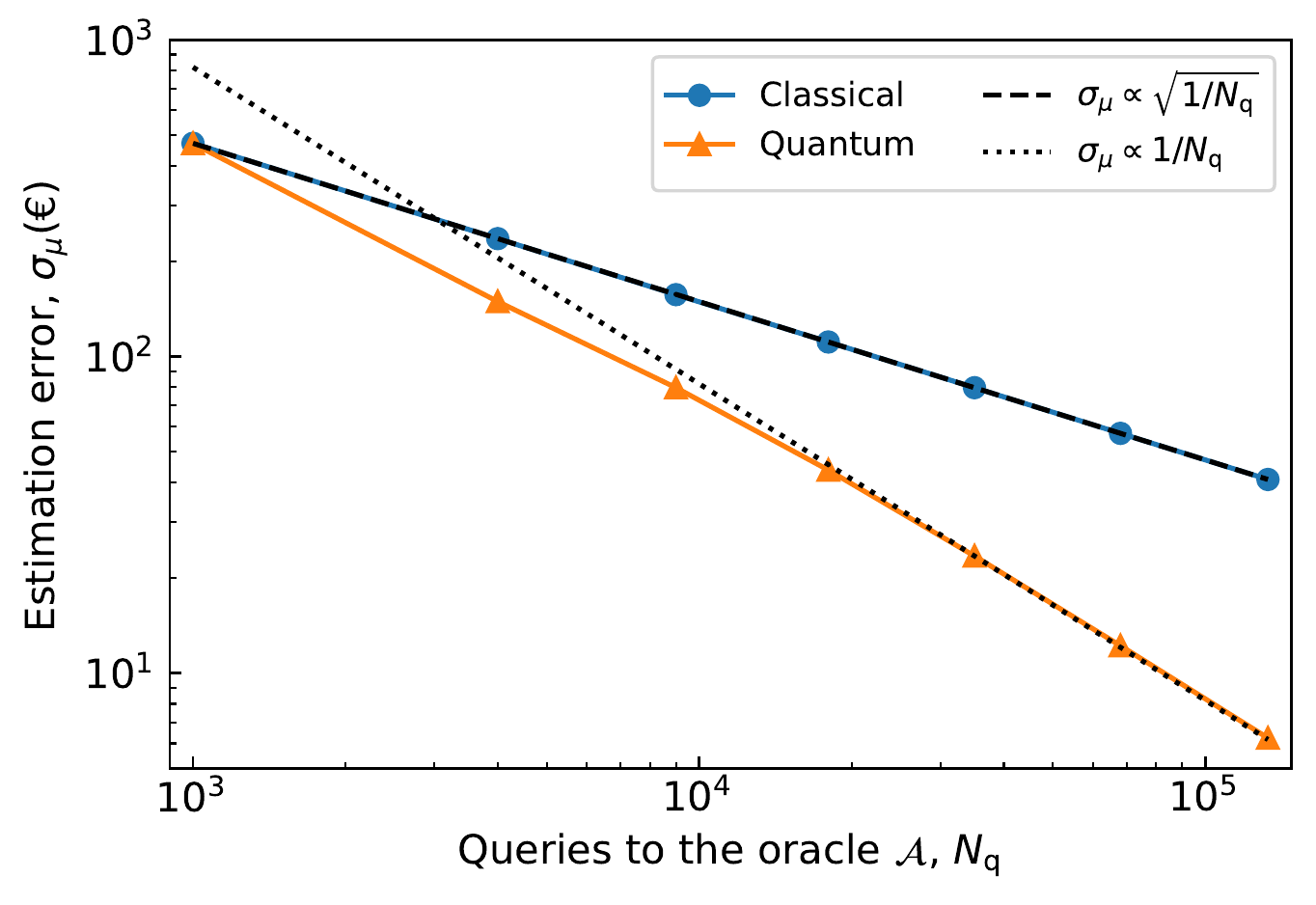}
  \caption{[Color online] Estimation error on the intrinsic long-term value for a stable market trend as a function of the number of queries to operator $\mathcal{A}$ (i.e., number of calls to the function to be estimated). We compare classical (blue dots) and ``quantum-enhanced" Monte Carlo (orange triangles) results. The black lines represent the large-$N_q$ rate of decrease in the error estimation for the quantum (dotted) and classical (dashed) cases.} 
  \label{fig:errors}
\end{figure}

The estimation error \footnote{Importantly, {\it error} here means the statistical error due to the intrinsic uncertainty in the sampling, either classical or quantum. It does {\it not} refer to the experimental error in the calculation, such as quantum gate errors. This experimental error is considered in the supplementary material.} on the mean intrinsic price can in theory be arbitrarily reduced by increasing the circuit depth in the quantum case, or increasing the number of samples in the classical case. In Fig.~\ref{fig:errors} we plot the estimation error versus number of calls $N_\text{q}$ to the operator $\mathcal{A}$, as computed by a simulator of the quantum hardware in the ``regular" scenario. In our algorithm, $N_\text{q}$ denotes the (classical) number of samples or (quantum) circuit depth and, therefore, it is proportional to the computational cost. The specific method we use to estimate the error in the forecast intrinsic price of the portfolio is explained in the Supplementary Material. The results in Fig.~\ref{fig:errors} clearly confirm a quadratic quantum speedup: whereas in classical Monte Carlo simulations the estimation error drops as $\sigma_{\mu} \propto N_{\rm{q}}^{-1/2}$, with quantum methods we can achieve $\sigma_{\mu} \propto N_{\rm{q}}^{-1}$, as shown in the figure. Or in financial words: {\it for the same computational cost, we estimate the intrinsic portfolio value with a quadratically-lower risk.}

\bigskip 
\emph{Conclusions.-} In this paper we have presented a method for efficiently estimating the intrinsic long-term price of a portfolio of assets using quantum computation. This method relies on Quantum Amplitude Estimation to evaluate the mean of a modified Gordon-Shapiro model. We demonstrate the effectiveness of our algorithms by estimating the intrinsic long-term value of a portfolio of 5 assets on present day quantum computers, namely, those provided by IonQ and AQTION, both of them based on trapped ions. Our quantum model outputs results which are consistent with the classical Monte Carlo benchmark, but with quadratically smaller statistical error for the same computational cost. 

In order to price such a large portfolio on present-day quantum computers, we had to make a number of assumptions about the nature of the stochastic variables. We anticipate that, in the near future, quantum computers will become available with a higher quantum volume. This will allow to represent more stochastic variables also in a more accurate way, in turn allowing for even more accurate calculations of portfolio prices. Another interesting direction for future work would involve finding implementations of operator $\mathcal{W}$ that can optimally implement functions such as $\phi(x)=a/(bx+c)$. This would allow us to price our portfolio without linearising Eq.~\eqref{eq:gordon_shapiro_full}, leading to an overall improvement of the result's accuracy.

\bigskip 

{\bf Acknowledgments:} We gratefully acknowledge funding from the EU H2020-FETFLAG-2018-03 under Grant Agreement No. 820495. We also acknowledge support by the Austrian Science Fund (FWF), through the SFB BeyondC (FWF Project No. F7109), as well as from Automatiq (FFG Project No. 872766), ELQO (FFG Project No. 884471), and the IQI GmbH. Moreover, we thank Christophe Jurczak, Pedro Luis Uriarte, Pedro Mu{\~n}oz-Baroja, Joseba Sagastigordia, Creative Destruction Lab, BIC-Gipuzkoa, DIPC, Ikerbasque, Basque Government and Diputaci\'on de Gipuzkoa for constant support, as well as Denise Ruffner for her special implication that guaranteed the success of this project. We also acknowledge the hard work and constant feedback from AQTION, IonQ, and Multiverse Computing fantastic teams.

\bibliography{bibliography}
\bibliographystyle{apsrev4-1}

\end{document}


\maketitle 

\section*{Quantum portfolio value forecasting: Supplementary Material}

In this supplementary material we provide details on the methods used to implement the quantum algorithm as well as to compute the estimation error of the calculations.

\subsection{Readout by Maximum Likelihood Estimation}

Maximum Likelihood Estimation (MLE) is a multi-circuit algorithm. It relies on the successive application of different powers of $\mathcal{Q}$, the Quantum Amplitude Amplification (QAA) operator, with $\mathcal{Q} = \mathcal{A} \mathcal{S}_0 \mathcal{A}^{\top} \mathcal{S}_{\chi}$ \cite{Suzuki2020,Brassard2002}. The operators $\mathcal{S}_{\chi}$ and $\mathcal{S}_0$ are defined such that: $\mathcal{S}_{\chi}\ket{x}_n \ket{1} = - \ket{x}_n \ket{1} \ \forall x$ and $\mathcal{S}_0\ket{0}_n\ket{0}=-\ket{0}_n\ket{0}$.

For the MLE method, we compute the likelihood function $\mathcal{L}$ describing each circuit's resulting measurements of the ancilla qubit after $n_{\rm{s}}$ measurements as follows:
\begin{equation}\label{eq:likelihood-func}
    \mathcal{L} = P(1)^{n_{\rm{g}}} [1-P(1)]^{n_{\rm{s}}-n_{\rm{g}}} \; .
\end{equation}
Here, $P(1)$ is the probability of measuring $1$ in the ancilla qubit and $n_{\rm{g}}$ is the number of $1$ measurements. For the classical case, that is, without QAA, $P(1)$ corresponds to Eq.(8) in the main paper. In the quantum enhanced circuits using QAA, that is, applying the $\mathcal{Q}$ operator one or more times, $P(1)$ is amplified \cite{Brassard2002, Montanaro2015}. Consequently, the error in the estimation of the mean value is reduced.

The next stage of the algorithm is to combine the likelihood functions for several circuits with different number of $\mathcal{Q}$ applications.
We thus obtain a single likelihood function describing the complete multi-circuit algorithm.
The MLE analysis consists of calculating the maximum of that resulting likelihood function to infer the most probable mean value.
The MLE algorithm does not require extra phase estimation operations after QAA \cite{kitaev1995quantum}. 
With the MLE algorithm, the quantum enhancement is thus still introduced using QAA, but depth circuit optimization is achieved due to the likelihood analysis.

In our paper, we combine circuits with different number of $\mathcal{Q}$ applications. We can define the vector $[m_0, m_1,..., m_{M-1}]$ in which each element correspond to the number of applications of $\mathcal{Q}$ per circuit, with a total of $M$ circuits. Since $\mathcal{Q}$ contains two $\mathcal{A}$ operators, this translates into
\begin{equation}
    N_{\rm{q}} = n_{\rm{s}} \sum_{k=0}^{M-1} (2 m_k + 1) \; 
\end{equation}
as the total number of queries to the oracle $\mathcal{A}$.
Fig.1 in the paper is calculated combining two circuits, i.e.: $M=2$. For the classical case, we use $[m_0 = 0, m_1 = 0]$, whereas for the quantum enhanced results we use $[m_0 = 0, m_1 = 2]$. On the other hand, Fig.2 compares the classical and quantum results as we increase the number of combined circuits, namely, $M=1, 2, 3, 4, 5, 6, 7$. For the classical results, $m_k = 0\ \forall k$, whereas for the quantum ones we use $[m_0 = 0, m_1 = 1, m_2 = 2, m_3 = 4, m_4 = 8, m_5 = 16, m_6 = 32]$.
In both figures, the number of measurements or shots per circuit is $n_{\rm{s}} = 1000$.

\subsection{Linearisation of $V$}

Following the recipe in Ref. \cite{Woerner2019}, we linearise the fair value in terms of the stochastic variables $\delta_{j, E}$ and $\delta_r$:
\begin{equation}
\label{eq:linearized_P}
V \approx \sum_{j=0}^{n_{\rm{a}}-1} \omega_j ( a_j + b_j \delta_{j, E} + c_j \delta_{r} + \mathcal{O}(\delta^2) ),
\end{equation}
with:
\begin{eqnarray}
    a_j & = & \frac{E_{j,1}}{1+r_{j,1}} + b_j  , \\
    b_j & = & \frac{E_{j,1}}{1+r_{j,1}} \frac{1+g_j}{1+r_{j,2}} \left( 1 + \frac{1+g_j}{\bar{r}_{\infty} + \nu_j - g_j} \right)  , \\
    c_j & = & - \frac{E_{j,1}}{1+r_{j,1}} \frac{(1+g_j)^2}{1+r_{j,2}} \frac{ \bar{r}_{\infty}}{(\bar{r}_{\infty} + \nu_j - g_j)^2}   .
\end{eqnarray}
Equation \eqref{eq:linearized_P} is a good approximation for $V$ when $\delta_{j, E}, \delta_r \ll 1$, which can be expected from a low volatility, mature market.

The intrinsic values $V_j$ can in general be any real positive number. We wish to encode this value to the amplitude of a quantum state (as explained in the main paper), which can take any value between 0 and 1. We therefore define the rescaled intrinsic value $\tilde{V}_j$ as:
\begin{equation}
\label{eq:scaling}
\tilde{V}_j = \frac{V_j - V_\text{min}}{V_\text{max} - V_\text{min}} \; .
\end{equation}
In the forthcoming sections of this Supplementary Material we specify the values of $V_\text{min}$ and $V_\text{max}$ used depending on the market trend analysed.

\subsection{Data loading}

For a portfolio composed of $n_a$ assets, the number of qubits necessary to estimate the portfolio's intrinsic value is:
\begin{equation}
n_\text{qubits} = q(n_a + 1) + 1,
\end{equation}
where $q$ is the bit depth of the $\delta_{j,E}$ and $\delta_{r}$.

In general, the variables $\delta_{j,E}$ are correlated random variables. As present day quantum computers have limited resources, we will assume in the following that $\delta_{j,E}=\delta_{E}$ is the same for all $j$. This is equivalent to assuming that all assets are subject to fluctuations from a random market to which they are heavily correlated. As such, the number of qubits needed is reduced to:
\begin{equation}
n_\text{qubits} = 2q + 1,
\end{equation}
In this paper, we use $q=2$, such that $n=4$ qubits are needed to define our problem, plus an extra ancilla qubit for the complete algorithm.

In general, $\mathcal{O}(2^n)$ quantum gates are needed to define $\mathcal{A}$ which loads an arbitrary distribution to a quantum state \cite{Mottonen2005}. We have observed that our distributions $\delta_E$ and $\delta_r$ are well approximated by a normal distribution. This is generally expected within an efficient market. As such, we are able to use the algorithm from Ref.\cite{Grover2002}, which efficiently loads a Gaussian distribution to a quantum state.

\subsection{Error calculation}

Each measurement of the ancilla qubit is a Bernoulli trial \cite{papoulis1984bernoulli}, i.e: a random variable with just two possible outcomes, namely, {\it good} or {\it bad} count (see Eq. 7 of the main paper). These have probabilities $P(0)$ and $P(1)$, respectively, which implies that the likelihood function used in the MLE analysis of each circuit's outcome has the form of Eq. \eqref{eq:likelihood-func}.
For a (classical) Bernoulli experiment with $n_{\rm{s}}$ measurements, the squared estimation error is as $\sigma_{\mu}^2 = P(1) [1 - P(1)] / n_{\rm{s}}$. 
For our algorithm, the estimation error is generalised as follows \cite{Suzuki2020}:
\begin{equation}
    \sigma_{\mu}^2 = \frac{P(1) [1 - P(1)]}{n_{\rm{s}} \sum_{k=1}^{M} (2 m_k + 1)^2} \; .
\end{equation}
That is, for the classical case, with $m_k = 0\ \forall k$, we recover $\sigma_{\mu} \propto n_{\rm{s}}^{1/2} = N_{\rm{q}}^{1/2}$. For the quantum case, however, the number of $\mathcal{Q}$ applications is $m_k \neq 0\ \forall k \neq 1$, which translates into a reduction in the estimation error. This implies we can understand QAA as an amplification of the total number of measurements of a circuit, which statistically reduces the error of the estimation.

\subsection{Input Data}

We study a portfolio of five indices. We invest 1000 euros in each asset, bought at the market value on the 2021-04-25. The corresponding number of shares held is presented in Table I in the paper. We summarize the input data purchased from a market data provider in Table \ref{tab:input_data}. We assume constant values for all the variables for the three different market trends analysed (bearish, stable, and bullish), except for the EPS growth rate.
\begin{table}[h]
\centering
\setlength{\tabcolsep}{3pt} 
\renewcommand{\arraystretch}{1.5} 
\begin{tabular}{| l || c c c c c|}
\hline
	\textbf{Index} & \textbf{SXXP} & \textbf{SPX} & \textbf{NKY} & \textbf{MXEF} & \textbf{EPRA} \\
	\hline
	$E_{j,1}$ (euros) & 23 & 171 & 1348.53 & 83 & 107 \\
	\hline
	$\nu_j$ (\%) & 5.12 & 4.73 & 4.50 & 7.50 & 5.12 \\
	\hline
	$r_1$ (\%) & 4.5 & 4.11 & 3.88 & 6.88 & 4.5 \\
	\hline
	$r_2$ (\%) & 4.92 & 4.53 & 4.30 & 7.30 & 4.92 \\
	\hline
	$r_{\infty}$ (\%) & 5.62 & 5.23 & 5 & 8 & 5.62 \\
	\hline
	$g_j$ (\%) - Bearish & -0.05 & -0.05 & -0.05 & -0.1 & -0.05 \\
	\hline
	$g_j$ (\%) - Stable & 1 & 1 & 1 & 1.5 & 1 \\
	\hline
	$g_j$ (\%) - Bullish & 1.8 & 1.8 & 1.8 & 1.8 & 1.8 \\
	\hline
\end{tabular}
\caption{Input data per index. }
\label{tab:input_data}
\end{table}

The long-term variables, $E_{j,2}$ and $r_{j,\infty}$, follow the distributions of the stochastic shifts $\delta_{j,E}$ and $\delta_r$. For these shifts, we assume bounded normal distributions characterised by $\mathcal{N} (\mu, \sigma)$, with $\mu$ the mean value and $\sigma$ the standard deviation. Namely, we use $\delta_{j,E} = \delta_{E} = \mathcal{N} (0, 0.1)$ and $\delta_{r} = \mathcal{N} (0, 0.01)$. In Table \ref{tab:deltas_bounds} we summarise the bounds used for the three different market trends studied in this paper. We assume bearish and bullish scenarios to be more volatile than stable ones, assuming the bearish market trend as the more volatile case.

\begin{table}[h]
\centering
\setlength{\tabcolsep}{8pt} 
\renewcommand{\arraystretch}{1.5} 
\begin{tabular}{| l|| c c c|}
\hline
      \textbf{Market} & \textbf{Bearish} & \textbf{Stable} & \textbf{Bullish}\\
      \hline
      {$\delta_E$} & $\pm 0.5$ & $\pm 0.3$ & $\pm 0.4$ \\
      \hline
      {$\delta_r$} & $\pm 0.05$ & $\pm 0.03$ & $\pm 0.04$ \\
      \hline
\end{tabular}
\caption{Bounds used for the distributions of the stochastic shifts per market trend.}
\label{tab:deltas_bounds}
\end{table}

Substituting the input data of Table \ref{tab:input_data} and using the bound values of Table \ref{tab:deltas_bounds} for the stochastic variables in Eq. \eqref{eq:linearized_P}, we can determine $V_{\rm{min}}$ and $V_{\rm{max}}$ values for Eq. \eqref{eq:scaling}. Table \ref{tab:V_min_max} shows $V_{\rm{min}}$ and $V_{\rm{max}}$ for the three market trends analysed.
\begin{table}[h]
\centering
\setlength{\tabcolsep}{8pt} 
\renewcommand{\arraystretch}{1.5} 
\begin{tabular}{| l|| c c c |}
\hline
      \textbf{Market} & \textbf{Bearish} & \textbf{Stable} & \textbf{Bullish}\\
      \hline
      {$V_{\rm{min}}$ (euros)} & $2570.89$ & $3666.60$ & $3757.44$ \\
      \hline
      {$V_{\rm{max}}$ (euros)} & $5735.25$ & $6656.09$ & $8561.35$ \\
      \hline
\end{tabular}
\caption{Bounds used for the distributions of the stochastic shifts per market trend.}
\label{tab:V_min_max}
\end{table}

\subsection{Circuit errors}

For the AQTION ion trap we have estimated the circuit errors by measuring the Hellinger fidelities $\mu$ between the expected probability distribution $p_i$ (obtained from simulations) and the estimated one $q_i$. This is given by 
\begin{equation}
\mu \equiv \sum_i \sqrt{p_i q_i}. 
\end{equation}
We have measured this quantity for two circuits: (i) for state preparation, and (ii) for state preparation and quantum amplitude estimation (QAE). In both cases we have considered the bearish, regular and bullish scenarios. Results are summarized in Tables \ref{tab:hel1} and \ref{tab:hel2}, where we also show the dispersion $\sigma$ of the fidelity. More specifically, in Table \ref{tab:hel1} we show the fidelities with respect to the QPU, and in Table \ref{tab:hel2} with respect to the QPU combined with postselection (where we use information from the simulation to make a ``wise" postselection of the results from the QPU). As we can see, the fidelities are not so good for the case of state preparation + QAE if we just take the results from the QPU but, however, this improves very significantly if we include postselection. A similar benchmark for the IonQ processor is currently work in progress. 

\begin{table}[h]
\centering
\setlength{\tabcolsep}{8pt} 
\renewcommand{\arraystretch}{1.5} 
\begin{tabular}{| c || c | c | c |}
\hline
     ~\textbf{Circuit}~ & ~\textbf{Bearish}~ & ~\textbf{Regular}~ & ~\textbf{Bullish}~ \\
     \hline 
     ~State preparation~ & ~$\mu \approx 0.859, \sigma \approx 0.000899$~ & ~$\mu \approx 0.903, \sigma \approx 0.000796$~ & ~$\mu \approx 0.914, \sigma \approx 0.000505$~ \\   
  ~State preparation + QAE~ & ~$\mu \approx 0.361, \sigma \approx 0.000709$~ & ~$\mu \approx 0.593, \sigma \approx 0.000602$~ & ~$\mu \approx 0.546, \sigma \approx 0.000456$~ \\  
      \hline
\end{tabular}
\caption{Hellinger fidelities and associated dispersions with respect to the QPU outcome.}
\label{tab:hel1}
\end{table}

\begin{table}[h]
\centering
\setlength{\tabcolsep}{8pt} 
\renewcommand{\arraystretch}{1.5} 
\begin{tabular}{| c || c | c | c |}
\hline
     ~\textbf{Circuit}~ & ~\textbf{Bearish}~ & ~\textbf{Regular}~ & ~\textbf{Bullish}~ \\
     \hline 
     ~State preparation~ & ~$\mu \approx 0.995, \sigma \approx 0.000428$~ & ~$\mu \approx 0.981, \sigma \approx 0.000865$~ & ~$\mu \approx 0.992, \sigma \approx 0.000548$~ \\   
  ~State preparation + QAE~ & ~$\mu \approx 0.980, \sigma \approx 0.000760$~ & ~$\mu \approx 0.972, \sigma \approx 0.000985$~ & ~$\mu \approx 0.980, \sigma \approx 0.000818$~ \\  
      \hline
\end{tabular}
\caption{Hellinger fidelities and associated dispersions with respect to the QPU outcome + postselection.}
\label{tab:hel2}
\end{table}

\bibliography{bibliography}
\bibliographystyle{apsrev4-1}